\documentstyle[psfig]{elsart}
\begin{document}
\begin{frontmatter}
\title{Ising thin films with modulations and surface defects}

\author[AC]{W. Selke\thanksref{ca}},
\author[ER,NA]{M. Pleimling},
\author[BE]{I. Peschel},
\author[BE]{M. Kaulke},
\author[BE]{M.-C. Chung} and
\author[AC]{D. Catrein}
\address[AC]{Institut f\"ur Theoretische Physik, Technische Hochschule, D--52056 Aachen, Germany}
\address[ER]{Institut f\"ur Theoretische Physik 1, Universit\"at Erlangen-N\"urnberg, D--91058 Erlangen, Germany}
\address[NA]{Laboratoire de Physique des Mat\'eriaux, Universit\'e Henri Poincar\'e Nancy I, B.P. 239, F--54506 Vand{\oe}uvre l\`es Nancy Cedex, France}
\address[BE]{Fachbereich Physik, Freie Universit\"at Berlin, D--14195 Berlin, Germany}
\thanks[ca]{Corresponding author:
tel: +49-241-807029; fax: +49-241-8888188;
e-mail: selke@physik.rwth-aachen.de}
 
\begin{abstract}
Properties of magnetic films are studied in the framework
of Ising models. In particular, we discuss critical phenomena of 
ferromagnetic Ising films with straight lines of
magnetic adatoms and straight steps on
the surface as well as phase diagrams of the axial
next--nearest neighbour Ising (ANNNI) model for thin films
exhibiting various spatially modulated phases.
\end{abstract}

\end{frontmatter}

\newpage

Magnetism in thin films has attracted much interest, both
experimentally and theoretically \cite{c1,c2}. In this paper, we
shall deal, in the framework of Ising models, with critical
properties of films having
different types of surface defects as well as with phase diagrams
of thin films displaying spatially modulated magnetic
structures, extending our recent work \cite{c3,c4}.

To analyse the role of surface imperfections at criticality, we consider
the ferromagnetic nearest--neighbour Ising model
on the simple cubic lattice for films of $L$ layers, using
Monte Carlo and the density--matrix renormalization group
techniques. To be specific, we study films with a straight line of magnetic
adatoms or a straight
step of monoatomic height. Both types of defects, lines
and steps, are assumed to run across the entire surface layer
through its center and along one
of the crystallographic axes. The magnetic couplings at the surface, $J_s$,
may be identical or different from those in the bulk, $J_b$. The local
interactions at the defect may
be modified, e.g., the couplings between neighbouring spins
in the additional line \cite{c3}.

Let us first discuss results for films with lines of magnetic
adatoms (our work has been motivated by related
experiments \cite{c5}, albeit there the lines of adatoms are not
coupled magnetically to the substrate).-- The temperature, at which
the transition between the ferromagnetic and the paramagnetic
phases occurs, rises with the film
thickness, $T_c= T_c(L)$, as shown in Fig.1. Of course, the 
magnetization in the additional line, $m_l$, is usually different
from the surface
magnetization, see also Fig. 1, because
of the reduced coordination number and the, possibly, modified local
couplings.

\begin{figure}
\centerline{\psfig{figure=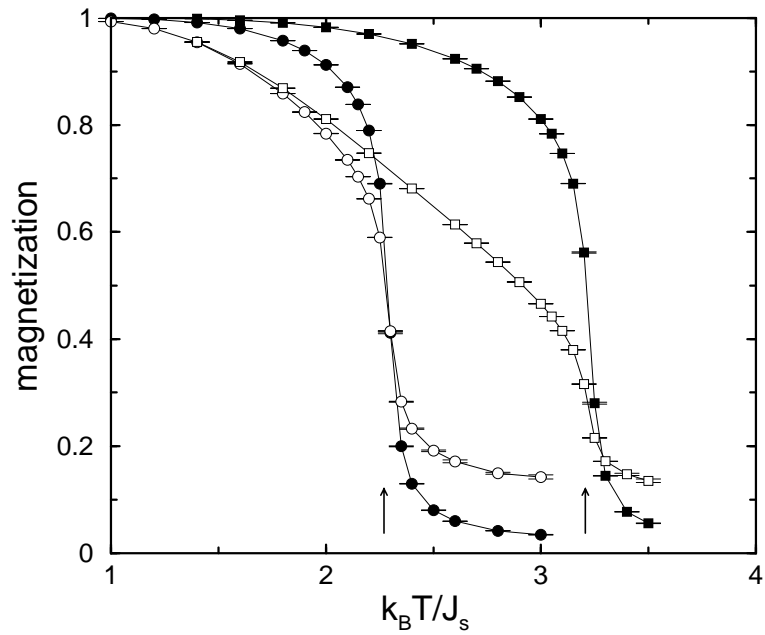,angle=270,width=10cm}} 
\caption{Simulated magnetization in the line of adatoms (open symbols)
and of the perfect surface (full symbols) of Ising models
with equal couplings for $L= 1$ (circles) and
2 (squares); each layer has $81 \times 80$ spins.}  
\label{fig1} \end{figure}

\begin{figure}
\centerline{\psfig{figure=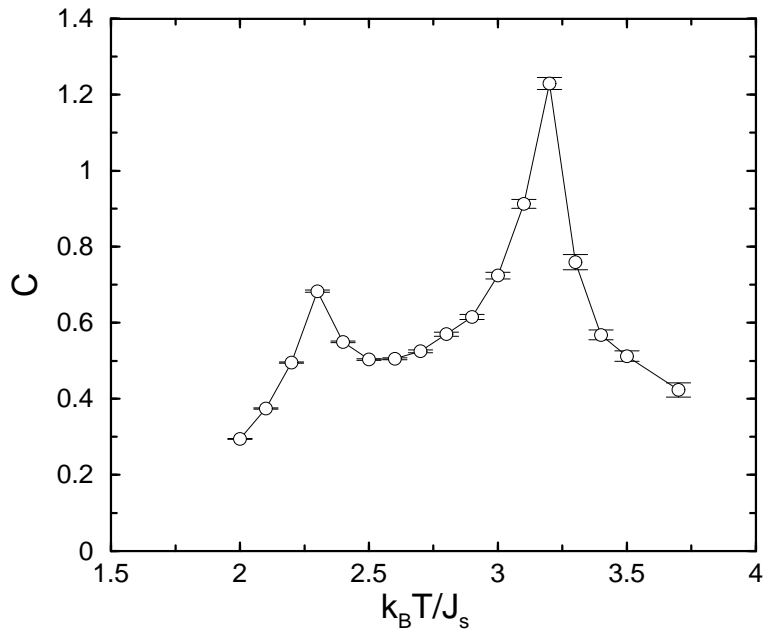,angle=270,width=10cm}} 
\caption{Simulated specific heat $C$ of the Ising model with
equal couplings and with a step dividing the
system into two halfs of one and two layers; the bottom layer
consists of $120 \times 120$ spins.}  
\label{fig2} \end{figure}
  
To elucidate critical properties at the line as compared to those
of the perfect surface and of the bulk, one may monitor the 
effective exponent $\beta_{eff}$. That quantity follows from the
ansatz for the magnetization $m \propto t^{\beta_{eff}}$, where
$t$ is the reduced temperature $|T_c(L)-T|/T_c(L)$. For
$t \longrightarrow 0$, $\beta_{eff}$ becomes the true critical
exponent $\beta$.-- In accordance with
general considerations on critical phenomena
at defect lines in systems with two--dimensional critical
fluctuations \cite{c6}, we find the critical exponent $\beta_l$
of the magnetization in the line of adatoms to be non--universal. In
films of finite thickness $L$, it varies from 0 to 1/2, decreasing
when strengthening the local couplings at the line, and
depending only rather weakly on $L$ \cite{c3}. In contrast, the
critical exponent of the magnetization of a perfect surface
is 1/8 for all finite values of $L$. It remains to be 1/8 
even in the thermodynamic limit, $L \longrightarrow \infty$, when
the surface orders at a higher temperature than the bulk ('surface
transition') due to sufficiently strong interactions in the
surface, $J_s$. When the surface and the bulk order at the same
temperature ('ordinary transition'), the critical exponent of the
surface magnetization is, in the limit $L \longrightarrow \infty$, about
0.80; this quite large
value leads, for finite $L$, to a pronounced increase of
the effective exponent away from criticality, both for the
magnetization of the surface and in the line \cite{c3}. That behaviour
should be taken into account in interpreting
properly, e.g., experimental data.

Another type of surface defect is the straight step of monoatomic
height, dividing the film in two halfs of thickness $L$ and
$L+1$. Accordingly, we are dealing with a composite system displaying
two distinct phase transitions at $T_c(L)$ and $T_c(L+1)$ \cite{c3}, as
signalled, e.g., by the two--peak structure of the specific heat, see
Fig. 2 (attention may be drawn to related recent
experimental observations \cite{c7}). Because $T_c(L+1) > T_c(L)$, the
magnetization at the step edge is expected to vanish
on approach to $T_c(L+1)$ similarly to the magnetization at the surface of
a two--dimensional Ising model. Thence, the critical exponent ist
expected to be 1/2 \cite{c6}, in accordance with our simulational
findings \cite{c3}.

Secondly, we study the influence of the film thickness on spatially
modulated magnetic phases in the framework of the
ANNNI model \cite{c8} on
the simple cubic lattice with $L$ layers \cite{c4}. Neighbouring spins
in each of the $L$ layers are coupled
ferromagnetically, $J_0 > 0$. Perpendicular to the layers
ferromagnetic interactions, $J_1 > 0$, between
neighbouring spins in adjacent layers, compete with
antiferromagnetic couplings, $J_2 < 0$, between axial next--nearest neighbour
spins. Our work extends a recent analysis on  
the influence of the film surface on critical
properties of the Lifshitz point \cite{c9}.

\begin{figure}
\centerline{\psfig{figure=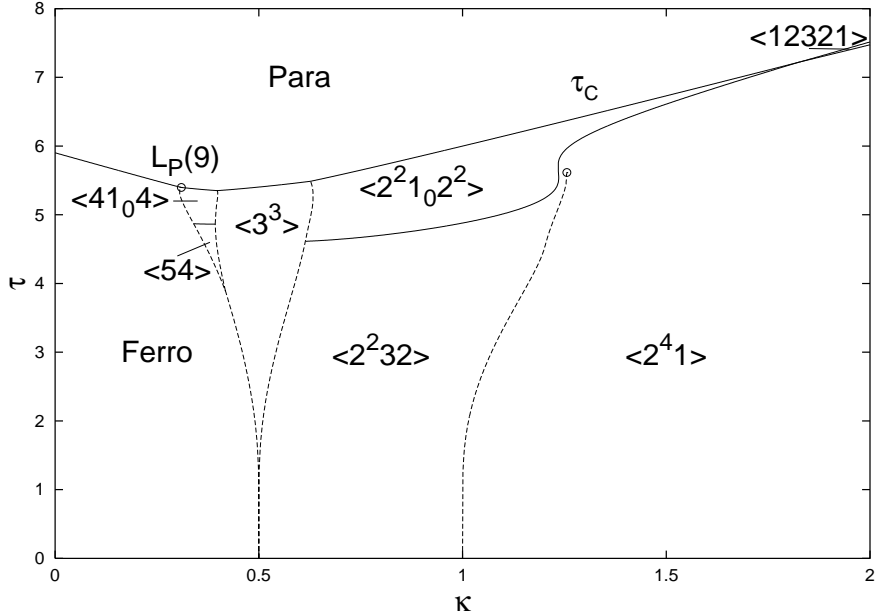,width=12cm}} 
\caption{Phase diagram of the ANNNI model with 9 layers, in
mean--field approximation; $\tau= k_BT/J_0$. Phases are denoted
in the usual manner, Refs. 4 and 8.}  
\label{fig3} \end{figure}

\begin{figure}
\centerline{\psfig{figure=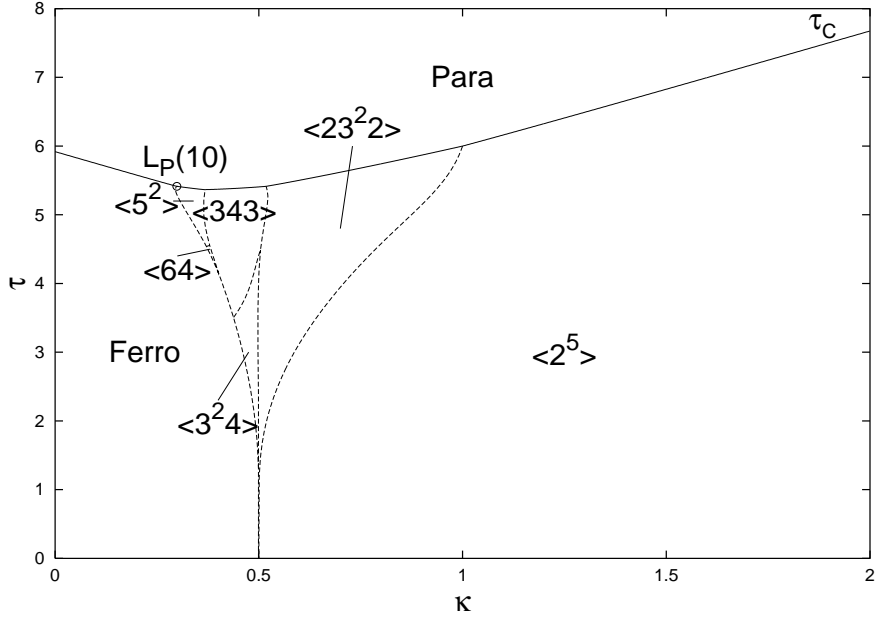,width=12cm}} 
\caption{Phase diagram of the ANNNI model with 10 layers, in mean--field
approximation.}
\label{fig4} \end{figure}
  
For each film thickness $L$, we determine, using mean--field
theory, Monte Carlo simulations and low--temperature series, a 
distinct phase diagram (setting $J_0= J_1$), in the
($\kappa = -J_2/J_1$, $k_BT/J_0$) plane, consisting of
various modulated phases. The cases $L=5$, 7, and 8 have been
depicted before \cite{c4}. In Figs. 3 and 4, we show the mean--field
phase diagrams for $L= 9$ and 10. Compared to the ANNNI model in
the thermodynamic limit, $L \longrightarrow \infty$, main differences
due to the finite film thickness are: (i) The ordered phases
occurring directly below the transition line to the paramagnetic
phase are either symmetric or antisymmetric about the center of the
film. In fact, neighbouring phases have different parity, with their
number of extrema in the layer magnetization
$m(z), z= 1, 2,... L$, increasing by one when
increasing $\kappa$. (ii) For $L$ odd, the center layer of the 
antisymmetric variants of those phases is disordered ('partially disordered
phases'). (iii) For $L$ odd, a transition line arises from
$(\kappa= 1, T= 0)$, in addition to the multiphase point at
$(\kappa= 1/2, T= 0)$ being present at all $L$. The phases springing
from the multiphase point may include structures (having one 4--band
or one 5--band of layers with equally oriented spins) being not stable
in the thermodynamic limit.

Of course, it is of interest to study the robustness of our findings
against modifications of the model parameters. In any event, results
may provide some guidance to explain possible experimental 
realizations for thin films with magnetic superstructures (here, attention
is drawn to recent local--spin--density calculations of thin $Fe$ films
on $Co$ substrates showing distinct and complex phase
diagram for various film thicknesses \cite{c10}).

\end{document}